\documentclass[pra,amsmath,amssymb,aps,showpacs,10pt,superscriptaddress,twocolumn]{revtex4-1}
\usepackage[utf8]{inputenc}
\usepackage{amsmath}
\usepackage{latexsym}
\usepackage{amsfonts}
\usepackage{epsfig}
\usepackage{psfrag}
\usepackage{float}
\usepackage{lipsum}
\usepackage{graphicx}
\usepackage{hyperref}
\usepackage{fixltx2e}
\usepackage{CJK}
\usepackage{amssymb,amsmath}
\usepackage{color}
\usepackage{soul,xcolor}
\usepackage[normalem]{ulem}

\newcommand{\bea}{\begin{eqnarray}}
\newcommand{\ea}{\end{eqnarray}}
\newcommand{\eea}{\end{eqnarray}}

\newcommand{\sumint}[1]
{\begin{array}{c} \\
{{\textstyle\sum}\hspace{-1.1em}{\displaystyle\int}}\\
{\scriptstyle{#1}}
\end{array}}

\usepackage[percent]{overpic}

\begin{document}

\preprint{AIP/123-QED}

\title{Crystallographic orientation dependence of work function:\\ Carbon adsorption on Au surfaces}

\author{H.Z. Jooya}
\email{hzjooya@cfa.harvard.edu}
\affiliation{%
ITAMP, Harvard-Smithsonian Center for Astrophysics, Cambridge, Massachusetts 02138, USA
}%

\author{X. Fan ({\begin{CJK*}{UTF8}{gbsn}星辰\end{CJK*}})}
\affiliation{%
ITAMP, Harvard-Smithsonian Center for Astrophysics, Cambridge, Massachusetts 02138, USA
}%

\author{K.S. McKay}
 \affiliation{%
National Institute of Standards and Technology, 325 Broadway, Boulder, Colorado 80305, USA
}%

\author{D.P. Pappas}
\affiliation{%
National Institute of Standards and Technology, 325 Broadway, Boulder, Colorado 80305, USA
}%

\author{D.A. Hite}
\affiliation{%
National Institute of Standards and Technology, 325 Broadway, Boulder, Colorado 80305, USA
}%

\author{H.R. Sadeghpour}
\affiliation{%
ITAMP, Harvard-Smithsonian Center for Astrophysics, Cambridge, Massachusetts 02138, USA
}%


\begin{abstract}
We investigate the work function (WF) variation of different Au crystallographic surface orientations with carbon atom adsorption. \textit{Ab-initio} calculations within density-functional theory are performed on carbon deposited (100), (110), and (111) gold surfaces. The WF behavior with carbon coverage for the different surface orientations is explained by the resultant electron charge density distributions. The dynamics of carbon adsorption at sub-to-one- monolayer (ML) coverage depends on the landscape of the potential energy surfaces. At higher ML coverage, because of adsorption saturation, the WF will have weak surface orientation dependence. This systematic study has consequential bearing on studies of electric-field noise emanating from polycrystalline gold ion-trap electrodes that have been largely employed in microfabricated electrodes. 

%
\end{abstract}

\keywords{Suggested keywords}
\maketitle
The work function (WF) of a metal is defined as the minimum energy needed to remove an electron from the bulk to a point far from the metal surface. The local WF of a surface, which can be experimentally measured using Kelvin-probe microscopy\cite{JOOYA2018}, is affected by the electronic distribution of charge on the surface. When atoms or molecules are adsorbed on a surface, charge transfer between adsorbates and the surface causes changes to the local dipole moment and the distribution of charge on the surface; consequently the WF changes. Surface diffusion of adsorbed species causes fluctuations in the WF and has been identified as a possible source of the electric-field noise \cite{Safavi2011,Safavi2013,Kim2017} detected in trapped ion experiments \cite{Hite2012,Hite2013}. Electric-field noise from surfaces is a major barrier to trapped-ion quantum computing, and an understanding of the underlying physical mechanisms is important for the elimination of this obstacle \cite{Brownnutt2015}.

Theoretical efforts to quantify the electric-field noise resulting from adsorbate diffusion depend on the magnitude of the dipole moment induced by the adsorbed adatom on the electrode surface. The magnitude of the dipole moment, $\Delta \mu$, is directly related to the surface WF, $\Delta W$, by $\Delta \mu=\Delta W \epsilon_{0} A/e$, where $\epsilon_0$ is the electric permittivity of free space, $A$ is the surface area occupied by one adatom, and $e$ is the electron charge.  Experimental measurements of $\Delta W$ can be used to directly measure the magnitude of the dipole moment induced by adsorbates.  

Gold is commonly used for ion-trap electrodes due to its noble characteristics (resistance to corrosion and oxidation), and adventitious carbon is a ubiquitous contaminant for metal surfaces.  In Ref.\cite{JOOYA2018}, measurements of the WF of C on Au (110) surfaces were compared to density-functional theory (DFT) calculations to determine the magnitude of the local dipole moment as a function of coverage.  However, ion-trap electrodes are not composed of single crystals because they are often fabricated using evaporation or electroplating techniques, which result in a polycrystalline structure.  As the WF is known to depend on surface orientation, adsorption sites, and adsorbate coverage, an understanding of the dipole moment on polycrystalline Au surfaces is needed to refine theoretical estimates for adsorbate diffusion induced electric field noise in ion trap experiments.  In this work, the results of DFT calculations of the adsorbate dipole moment are used to describe the dependence of the WF on crystallographic orientation, namely the (100), (110), and (111) low-index faces.  

To investigate the crystalline orientation dependence of the WF, when carbon adatoms are adsorbed on gold surfaces, we performed DFT calculations using the Plane-Wave Self-Consistent Field (PWSCF) option in the QUANTUM-ESPRESSO (QE) distribution \cite{QE2009}. The local-density approximations (LDA) exchange-correlation functional with the Perdew and Zunger expression was employed. It has been shown that LDA DFT results more closely agree with experiment, because the LDA exchange energy contribution to the surface energy is overestimated, while the correlation contribution is significantly underestimated. These two errors nearly cancel each other\cite{PNAS2017}. For gold, the ultrasoft pseudopotential (USPP) was generated with the Rappe Rabe Kaxiras Joannopoulos (rrkjus) scheme with 11 valence electrons in a $5d^{10} 6s$ configuration. For carbon atoms, the $2s^2$ $2p^2$  electrons were treated explicitly as valence electrons in the Kohn-Sham (KS) equations, and the remaining cores were represented by USPP pseudopotentials. A 450 Ry ( 1 Ry=13.605698 eV) kinetic-energy cutoff for the charge density was applied. All structures were optimized with periodic boundary conditions applied using conjugate gradient method, accelerated using the Marzari-Vanderbilt smearing \cite{Marzari1999} with a width of 0.01 Ry. Structural optimization and property calculations were carried out using the Monkhorst-Pack special k-point scheme \cite{Monkhorst1976} with $6\times 6\times 1$ meshes for integration in the Brillouin zone of the slab systems. 

\begin{figure*}
\includegraphics[width=1.0\textwidth, height=8cm]{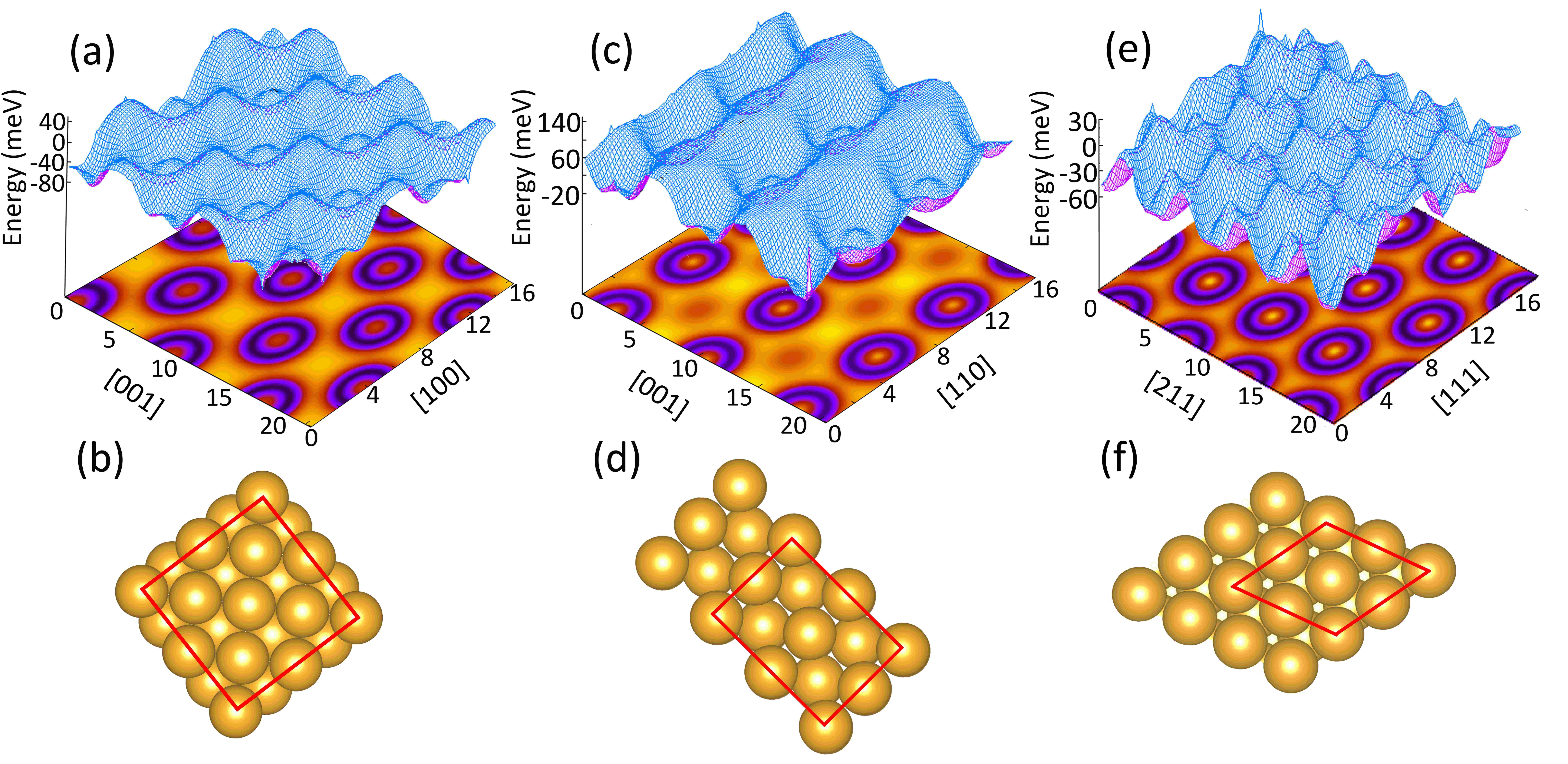}
\caption{\label{fig:Surfaces} Potential energy landscape (upper panel), and the top view of the corresponding crystallographic orientation surface (lower panel) of  Au(100) (a,b),  Au(110) (c,d), and  Au(111) (e,f). In the potential energy contour plot, the blue represents the minimum, and the red the maximum energy. The red boxes show the surface edges of the periodic boundary condition used in the DFT simulations of the adsorption processes.}
\end{figure*}

For each surface, the periodic supercell slab was constructed by cleaving relaxed bulk Au with lattice constant 4.14\text{\AA} , i.e., in good agreement with the experimental value of 4.0780\text{\AA} at 25$^{\circ}$ \cite{Dutta1963}. The slab model consisted of six-layer thick Au atoms with a normal ($1\times 1$) superstructure. The schematic of the three surfaces used in this study are shown in Fig.~\ref{fig:Surfaces} (lower panels). The top four layers were allowed to relax, while the bottom two layers were kept fixed to mimic the bulk. Although a large vacuum region (15\text{\AA}) was used between periodic slabs, the creation of dipoles upon adsorption of atoms on only one side of the slab can lead to spurious interactions between the dipoles of successive slabs. In order to circumvent this problem, a dipole correction was applied by means of a dipole layer placed in the vacuum region following the method outlined by Bengtsson \cite{Bengtsson1999}. As demonstrated in our previous work \cite{Safavi2013,Kim2017,JOOYA2018}, the introduction of this artificial dipole layer in the vacuum region does not modify the local potential near the surface where adsorption occurs. The adsorption energy of carbon on each surface is calculated as $E_{ads}=E_{S+C}-E_{S}-E_{C}$, where $E_{S+C}$, $E_{S}$, and $E_C$ are the total energies of the surface with a carbon adatom, of the bare surface. and of an isolated C atom, respectively.

\begin{figure}
\includegraphics[width=1.0\linewidth, height=9cm]{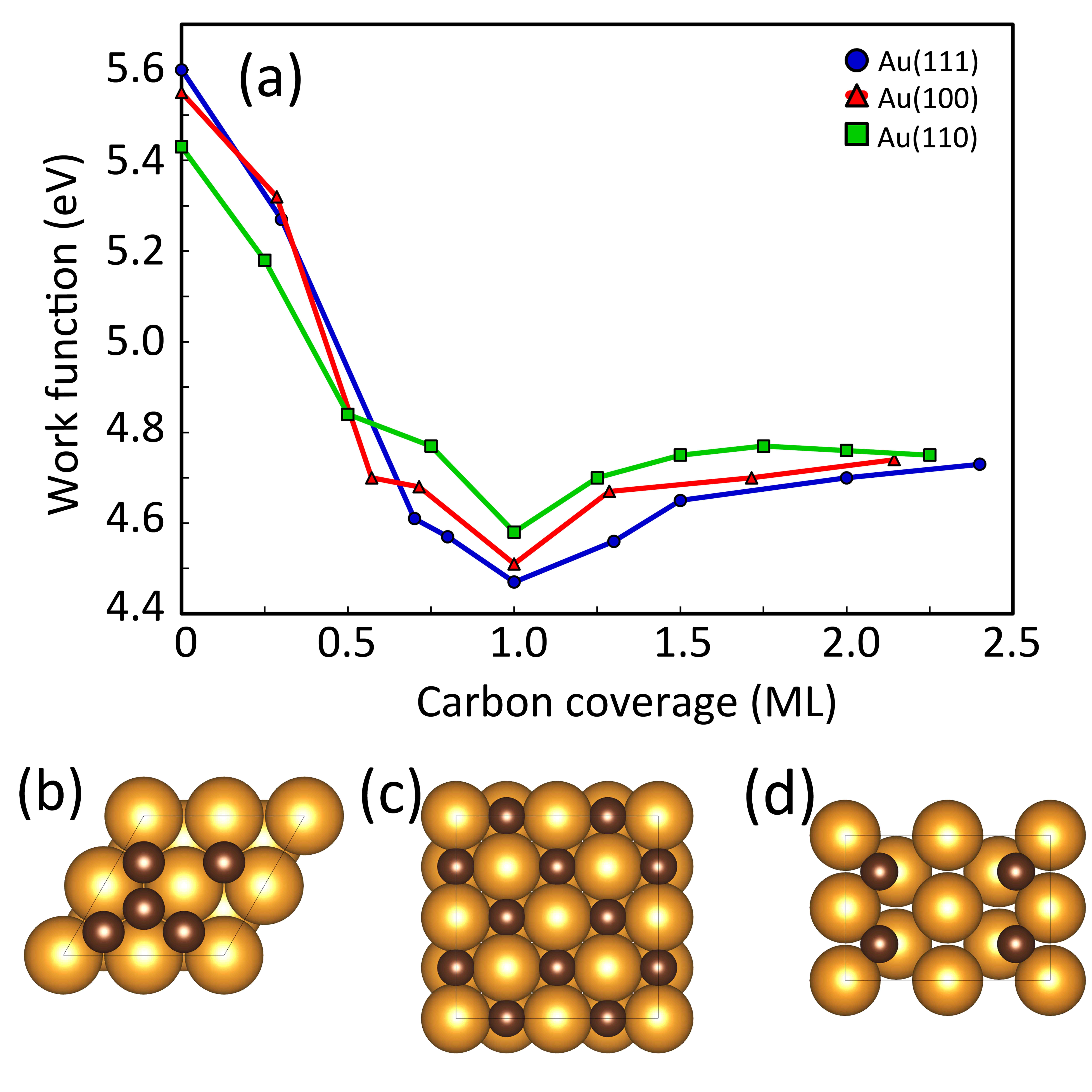}
\caption{\label{fig:Work function} (a) WF versus carbon coverage for different gold orientation surfaces. The rate at which the WF approaches the graphite value after 1 ML coverage is the highest for Au(110), followed by Au(100), and then Au(111). The data points are connected to improve the readability of the plots. The top view of the carbon covered (b) Au(111), (c) Au(100), and (a) Au(110) surfaces.}
\end{figure}

Upper panels in Fig.~\ref{fig:Surfaces} show the potential energy landscapes that in which a carbon atom is exposed, when it approaches each of the gold surfaces (lower panels). The potential energy surface (PES) is obtained by post processing the SCF results of the relaxed surface using the QE package. The adsorption energy of a fourfold coordinated C at the bridge site on Au(100) is 5.19 eV with bond length of 2.30\text{\AA}. The pseudothreefold site in Au(110) surface has an adsorption energy of 5.43 eV with bond length of 2.06\text{\AA} from the adjacent gold atoms. The threefold fcc site on Au(111) gives 4.54 eV to adsorb a carbon atom at 3.08\text{\AA}. These adsorption sites are illustrated in the lower panels in Fig.~\ref{fig:Work function}. In each case, the position of the carbon atoms can be determined by analyzing the PES. These positions can also be confirmed by \textit{ab-initio} molecular dynamics simulations \cite{JOOYA2018}. The preferred adsorption of carbon adatoms on fcc sites of Au(111) supports the formation of a graphene monolayer at higher coverage. This is in agreement with previous theoretical studies \cite{Santiago2014,Khomyakov2009}, and scanning tunneling microscopy of graphene nanoflakes on Au(111) \cite{Tesch2016}. Monolayer graphene has also been synthesized on hex-reconstruction of Au(100) \cite{Zhou2016}, and reported as nanoribbon growth on Au(110) \cite{Della2016}.  

The simplest model to predict how WFs may change due to the adsorption of atomic species was proposed by Topping. This model predicts that the degree of electron transfer from the adsorbed atom to the metal surface is proportional to the difference in electronegativities of the adsorbate and the substrate. Carbon atoms are only slightly more electronegative than the gold substrate (2.55 for C, and 2.54 for Au). Therefore the deposited carbon adatoms on the gold surface would be slightly polarized negatively outward. This is also confirmed by the analysis of the core-level binding energies, which indicates that no significant shifts are measured for the Au 4\textit{p}, 4\textit{d}, and 4\textit{f} core levels due to the carbon adsorption \cite{JOOYA2018}. This is an indication of minimal charge transfer to the Au surface and correlates with the corresponding dipole moment change with increasing coverage \cite{Riffe1990}. 

\begin{figure*}
\includegraphics[width=0.8\linewidth, height=5cm]{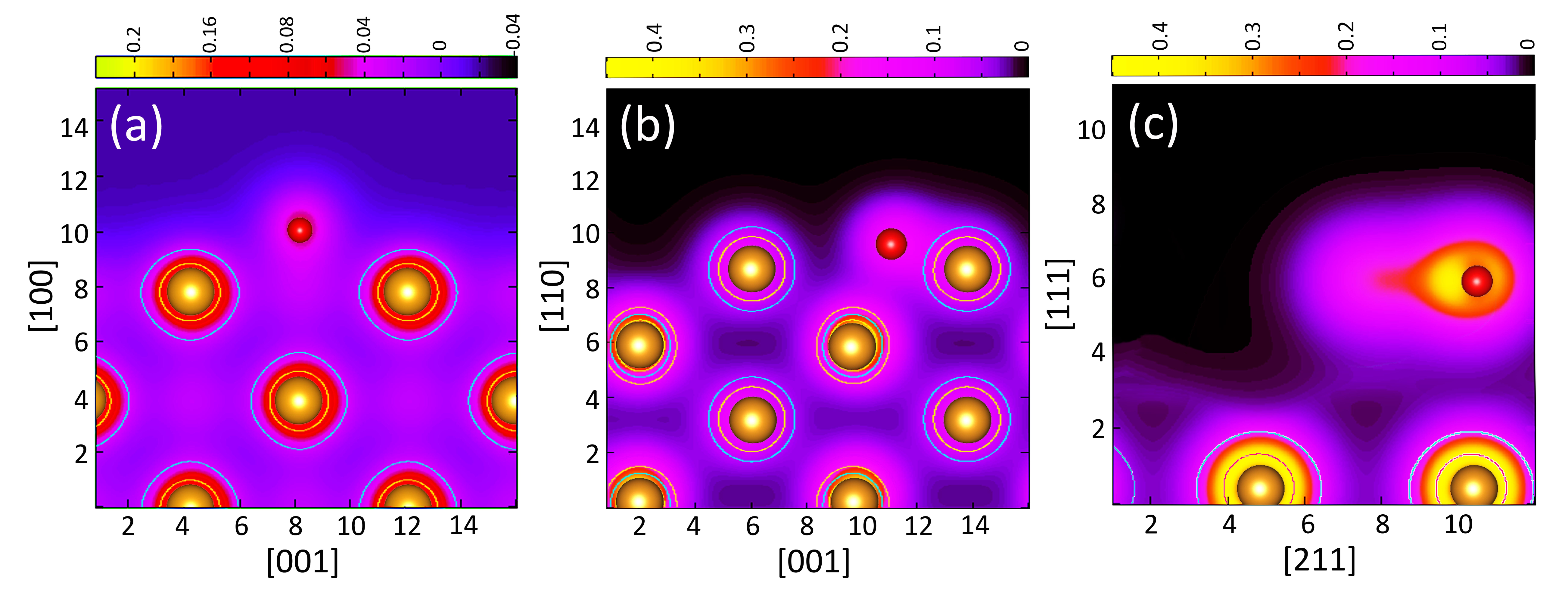}
\caption{\label{fig:Charge} Electron charge density distribution maps after carbons are adsorbed on (a) Au(100), (b) Au(110), and Au(111). The increase in charge corrugation lowers the WF giving rise to a dipole with the positive side outwards. This effect is minor for Au(110), more pronounced in Au(100), and distinctly noticeable for Au(111). The zoomed in view is provided in (c) to better illustrate the effect.}
\end{figure*}

According to the Topping model, with this minimal outward polarization, one would expect to detect an increase of the WF due to carbon adsorption. This is, however, in opposition to the behavior exhibited in Fig.~\ref{fig:Work function}. The WF of the clean surfaces are calculated to be 5.55 eV, 5.43 eV, and 5.60 eV for Au(100), Au(110), and Au(111), respectively. These values are in agreement with the previously reported LDA calculations \cite{Fall2000,PNAS2017,Singh-Miller2009}. As the top panel in Fig.~\ref{fig:Work function} shows, in each case starting from these values at zero coverage, the WF near linearly decreases with coverage. The WF then flattens off at monolayer coverage, toward minimum values of 4.51 eV, 4.58 eV, and 4.47 eV, for Au(100), Au(110), and Au(111), respectively.  After that, for all surfaces, the WF rises approximately to a value associated with the WF of the bulk adsorbate, \textit{i.e.} graphite. The dip in the value of the WF at 1 monolayer (ML) consistently appears for all of the surfaces we have considered. There is an indication that this is a universal behavior \cite{POLANSKI1973,Rhead1988,JOWETT1970}.

The correlation between the WF reduction for different gold surfaces is provided by analyzing the electron charge density distribution after carbon is adsorbed on Au(100), Au(110) and Au(111). The electron charge density plots presented in Fig.~\ref{fig:Charge} are obtained by post processing the SCF results of the relaxed surface. The effective role of the carbon adsorption on WF reduction is qualitatively explained by determining the change of the charge corrugation due to the adsorption process. This effect has previously been studied \cite{Rhead1988, JOOYA2018}. It is known \cite{Muscat1986,Serena1987} that any increase in this corrugation lowers the WF, giving rise to a dipole with the positive side outwards. 

As illustrated in Fig.~\ref{fig:Charge}(b), the carbon atom slightly penetrates into the potential well of the Au(110) surface (See Fig. ~\ref{fig:Surfaces}(c). Therefore the resultant corrugation of the electron charge density distribution due to the carbon adsorption is minor. This minimal effect is manifested as the smallest reduction in the WF at a 1 ML coverage of Au(110), 0.85 eV. 
As can be seen in Fig.~\ref{fig:Charge}(a), the effective role of the carbon adsorption on the electron charge density distribution is more pronounced for the Au(100) surface. This can also be understood by evaluating the potential energy landscape and the available adsorption sites for atomic carbons above the Au(100) surface (see Figs.~\ref{fig:Surfaces}(a,b)). The elevated charge corrugation in this case induces a 1.04 eV decrease in WF when 1 ML of carbon atoms is adsorbed on Au(100) (See Fig. ~\ref{fig:Work function} (a,b)). As shown in Fig. ~\ref{fig:Charge}(c) this effect is distinctly noticeable, due to the fact that fcc sites on the Au(111) surface assist the carbon interactions at closer distances, which in turn makes the carbon atoms relax at around 3.08 \text{\AA} from the surface. At submonolayer coverages, this makes the relatively smooth Au(111) surface more corrugated, which results in the largest drop of the WF at 1.13 eV. 

Scalability and stability of trapped ion qubits strongly depend on the electrode surface quality and the magnitude of the resulting electric-field noise. Accordingly, achieving high precision performance for such devices requires a thorough understanding of the electronic conditions on top of the electrode surface. Since in practice polycrystalline gold is commonly used for ion-trap electrodes, understanding the dependence of the surface WF with crystallographic orientation is of high importance. 

In this study we calculated, with \textit{ab-initio} density-functional theory, the WF dependence of different Au crystallographic surface orientations as a function of C adatom coverage. We analyzed the calculated behavior by the resultant electron charge density distributions. This study will provide guidance on the influence of electric-field noise emanating from polycrystalline ion-trap electrode surfaces.

\section{ Acknowledgments}
HZJ is supported by a theory grant from the NIST Physical Measurement Laboratory. HRS acknowledges support from the NSF through a grant for ITAMP at Harvard  University. XF conducted work as a high school intern in the summer of 2018 at ITAMP.  

\newpage
\bibliography{references}

\end{document}